\documentclass[12pt,onecolumn,draftcls]{IEEEtran}
\ifCLASSINFOpdf
   \usepackage[pdftex]{graphicx}
\else
   \usepackage[dvips]{graphicx}
\fi
\usepackage[cmex10]{amsmath}
\usepackage{url}
\usepackage{verbatim}

\usepackage{array}

\usepackage{bm}

\begin{document}
%
\title{A Theoretical Answer to ``Does the IRC-SINR of an Interference Rejection Combiner always Increase with an Increase in Number of Receive Antennas?"}



%
\author{\IEEEauthorblockN{Karthik Muralidhar}
}
 


\maketitle
\begin{abstract}
Interference rejection combiners (IRCs) are very popular in 4G/5G systems. In particular, they are often times used in co-ordinated multi-point (CoMP) networks where the antennas of a neighboring cell's base station (BS) are used in an IRC receiver, in conjunction with the antennas of the BS of a cell-edge UE's own cell, to improve the IRC-SINR  of a cell-edge user. But does the  IRC-SINR always increase with an increase in the number of antennas? In this paper, we attempt to answer the question theoretically. We give a theoretical derivation that quantifies the improvement in the IRC-SINR when the number of antennas increases by unity.  We show that this improvement in IRC-SINR is always greater than or equal to zero. Thus we prove that increasing the number of antennas even by unity will always improve the IRC-SINR. Selecting the extra antennas of the neighbouring cell can be viewed as a special case of antenna selection described in [1]. We also present the IRC-SINR improvement in an uplink CoMP scenario by simulations and verify that it indeed matches with the theoretical gains derived in this paper.

{\bf {\it Index Terms}} - Antenna selection, CoMP and IRC.
\end{abstract}
\section{Introduction}
IRCs are gaining popularity in 4G, 5G systems \cite{fieldtrial}, \cite{marsch} and \cite{ele}. They are used for cell-edge UEs in uplink co-ordinated multi-point (CoMP) framework, where the received signal at the antennas of the BS of two or more neighbouring cells is processed at a central processing unit (CPU). In \cite{antsel}, the authors dealt with IRCs in the context of antenna selection. In this paper, we show a theoretical derivation that as the number of antennas increase by unity, the IRC-SINR increase is always greater than or equal to zero. We show simulation results for IRC receivers in the context of uplink CoMP and show that the increase in IRC-SINR due to adding antennas of neighbouring cells to the IRC receiver completely matches the theoretical derivation done in this paper. While there could be many ways of doing the theoretical derivation, we start with and analyze in detail Equation (8) of \cite{antsel}, to derive the theoretical derivation. 

{\it Notation} : We follow Matlab notation to access parts of a vector or a matrix. Conjugate of $x$ is denoted by $x^*$. Real part of $x$ is denoted by $R\left\{x\right\}$. Identity matrix of appropriate dimension is denoted by ${\bf I}$.
\section{IRC-SINR gain due to increase in number of antennas}\label{snrgainsec}
In this section, let us assume that ${\bf h}_i,\: i=1,\ldots, Z$ are all channel vectors of $Z$ single-antenna UEs to $(N_R+1)$ receive antennas. The channel vectors are of dimension $(N_R+1)\times 1$. We will derive the IRC-SINR gain of the first UE as the number of antennas increase from $N_R$ to $N_R+1$.  Let
\begin{equation}\label{snrgain1}
\begin{array}{*{20}c}
   {\mathbf{h}_1  = \left[ {\begin{array}{*{20}c}
   {\mathbf{\bar h}_1 }  \\
   {\tilde{h}_1 }  \\

 \end{array} } \right],} & {\mathbf{P} = \left[ {\begin{array}{*{20}c}
   \mathbf{{\bar P}}  \\
   \bm{\rho}  \\

 \end{array} } \right] = \left[ {\begin{array}{*{20}c}
   {\mathbf{h}_2 } & {\mathbf{h}_3 } &  \cdots  & {\mathbf{h}_Z }  \\

 \end{array} } \right]}  \\

 \end{array} 
\end{equation}
where ${\bf{\bar h}}_1, {\bf {\bar P}}$ are the channel vector of the first UE and channel matrix of the next $(Z-1)$ UEs to the first $N_R$ antennas. The channel of the first UE to the $(N_R+1)$st antenna is $\tilde{h}_1$. Likewise the channel vector of the next $(Z-1)$ UEs to the $(N_R+1)$st antenna is $\bm \rho = \left[\tilde{h}_2, \ldots, \tilde{h}_{Z-1}\right]$. 

Let $\sigma^2$ be the variance of AWGN per antenna. Let us define 
\begin{equation}\label{snrgain2}
\begin{array}{*{20}c}
   \mathbf{A}  &=& \left( {\sigma ^2 {\bf I} + \mathbf{\bar P}^H \mathbf{\bar P}} \right)^{ - 1} ,  \\
   \mathbf{\bar h}_1^H \mathbf{\bar PA}{\bm\rho}^H  &=& y,  \\
   {\bm\rho} \mathbf{A} {\bm\rho}^H   &=& t  \\
\end{array} \cdot
\end{equation}
From Equation (8) in \cite{antsel}, the IRC-SINRs of an IRC receiver using $N_R$ and $N_{R+1}$ antennas are given by
\begin{equation}\label{snrgain3}
\begin{array}{*{20}c}
   \text{IRC-SINR}^{\left( {N_R } \right)}  &=& \frac{{\mathbf{\bar h}_1^H \mathbf{\bar h}_1 }}
{{\sigma ^2 } } - \frac{1}
{{\sigma ^2 } }\mathbf{\bar h}_1^H \mathbf{\bar P}\left( {\sigma ^2 {\bf I} + \mathbf{\bar P}^H \mathbf{\bar P}} \right)^{ - 1} \mathbf{\bar P}^H \mathbf{\bar h}_1 ,  \\
   \text{IRC-SINR}^{\left( {N_R  + 1} \right)}  &=& \frac{{\mathbf{h}_1^H \mathbf{h}_1 }}
{{\sigma ^2 }} - \frac{1}
{{\sigma ^2 }}\mathbf{h}_1^H \mathbf{P}\left( {\sigma ^2  {\bf I}+ \mathbf{P}^H \mathbf{P}} \right)^{ - 1} \mathbf{P}^H \mathbf{h}_1   \\
 \end{array} \cdot
\end{equation}
We also define 
\begin{equation}
   \mathbf{A}_{1}  = \left( {\sigma ^2 {\bf I} + \mathbf{ P}^H \mathbf{ P}} \right)^{ - 1}   \cdot
\end{equation}
Using Woodbury's identity \cite{woodbury} and \eqref{snrgain1}, we can write the recursive relation for $\mathbf{A}_{1}$ as
\begin{equation}\label{recursiveanr}
\mathbf{A}_{ 1}  = \mathbf{A}  - \frac{{\mathbf{A} \bm\rho ^H \bm\rho \mathbf{A} }}
{{1 + \bm\rho \mathbf{A} \bm\rho^H }}\cdot
\end{equation}
Define
\begin{equation}
\xi(N_R+1, N_R)={\text{IRC-SINR}}_\text{gain} = \text{IRC-SINR}^{\left( {N_R  + 1} \right)} - \text{IRC-SINR}^{\left( {N_R } \right)}\cdot
\end{equation}
From \eqref{snrgain3}, we have that
\begin{equation}\label{Omega}
\xi(N_R+1, N_R) = \frac{\left|\tilde{h}_1\right|^2}{\sigma^2} -\frac{\Omega}{\sigma^2}
\end{equation}
where
\begin{equation}\label{omegadef}
\Omega = \left\{\mathbf{h}_1^H \mathbf{P}\left( {\sigma ^2 {\bf I} + \mathbf{P}^H \mathbf{P}} \right)^{ - 1} \mathbf{P}^H \mathbf{h}_1 \right\} - \left\{\mathbf{\bar h}_1^H \mathbf{\bar P}\left( {\sigma ^2 {\bf I} + \mathbf{\bar P}^H \mathbf{\bar P}} \right)^{ - 1} \mathbf{\bar P}^H \mathbf{\bar h}_1\right\} \cdot
\end{equation}
From \eqref{snrgain1}, we have the relation ${\bf h}_1^H{\bf P} = {\bar{\bf h}}_1^H{\bar{\bf P}}+ \tilde{h}_1^*{{\bm \rho}}$. Using this relation and  the definition of ${\bf A}_1$, the first term in \eqref{omegadef} can be written as
\begin{equation}
\Omega _1  = \mathbf{h}_1^H \mathbf{P}\left( {\sigma ^2 {\bf I} + \mathbf{P}^H \mathbf{P}} \right)^{ - 1} \mathbf{P}^H \mathbf{h}_1 = \left( \bar{\bf h}_1^H{\bar{\bf P}}+ \tilde{h}_1^*{\bm \rho} \right)\mathbf{A}_1 \left( {\mathbf{\bar P}^H \mathbf{\bar h}_{  1}  + \tilde h_1 {\bm \rho} ^H } \right)\cdot
\end{equation}
Using \eqref{recursiveanr} we simplify $\Omega _1$ further as
\begin{equation}
\Omega _1  = \mathbf{\bar h}_1^H \mathbf{\bar PA\bar P}^H \mathbf{\bar h}_{  1}  - \frac{{\left| y \right|^2 }}
{{1 + t}} + 2R\left\{ {\tilde h_1^* \bm\rho \mathbf{A}_1 \mathbf{\bar P}^H \mathbf{\bar h}_{  1} } \right\} + \left| {\tilde h_1 } \right|^2{\bm\rho}\mathbf{A}_1 {\bm\rho} ^H  \cdot
\end{equation}
Substituting the above expression for $\Omega _1$ in \eqref{omegadef}, we arrive at
\begin{equation}
 \Omega= 
  2R\left\{ {\tilde h_1^* \bm \rho \mathbf{A}_1 \mathbf{\bar P}^H \mathbf{\bar h}_{  1} } \right\} + \left| {\tilde h_1 }^2 \right| \bm \rho \mathbf{A}_1 \bm \rho ^H  - \frac{{\left| y \right|^2 }}{{1 + t}} \cdot
\end{equation}
We now substitute \eqref{recursiveanr} in the above equation to simplify $\Omega$ further as 
\begin{equation}
\Omega=2R\left\{ {\tilde h_1^* y^* } \right\} + \left| {\tilde h_1 } \right|^2 t - \frac{{\left| y \right|^2 }}
{{1 + t}} - \frac{{\left| {\tilde h_1 } \right|^2 t^2 }}
{{1 + t}} - \frac{{2R\left\{ {\tilde h_1^* y^* t} \right\}}}
{{1 + t}}\cdot
\end{equation}
Using the above relation in \eqref{Omega}, we get
\begin{equation}\label{increase1}
\xi(N_R+1, N_R) = \frac{1}{\sigma^2}\left\{\left|{\tilde h}_1\right|^2 - 2R\left\{ {\tilde h_1^* y^* } \right\} - \left| {\tilde h_1 } \right|^2 t + \frac{{\left| y \right|^2 }}
{{1 + t}} + \frac{{\left| {\tilde h_1 } \right|^2 t^2 }}
{{1 + t}} + \frac{{2R\left\{ {\tilde h_1^* y^* t} \right\}}}
{{1 + t}}\right\}
\end{equation}
which gets simplified to
\begin{equation}\label{theoretical}
\xi(N_R+1, N_R) = \frac{\left| y \right|^2 + \left|{\tilde h}_1\right|^2 - 2R\left\{ {\tilde h_1^* y^* } \right\}}{\sigma ^2 \left( {1 + t} \right)} = \frac{{\left| {y - {\tilde h}_1^* } \right|^2 }}
{{\sigma ^2 \left( {1 + t} \right)}}
\end{equation}
which is always greater than or equal to zero. Hence proved.
The IRC-SINR gain as number of antennas increase by $a$ is 
\begin{equation}\label{sims}
\xi(N_R+a, N_R) =  \sum_{i=N_R}^{N_R+a-1} \xi(i+1, i)\cdot
\end{equation}

\section{Simulation results}
We consider four cells. Each cell has two UEs each with a single antenna. The BS in each cell has four antennas.  The single-cell IRC-SINR of an UE is obtained by applying the IRC algorithm to the signals received at the four antennas of the BS in the cell to which the UE belongs. The multi-cell IRC-SINR of an UE is obtained by applying the IRC algorithm to all 16 antennas of the combined BSs of all four cells (this happens in the CPU of cloud radio access network). Let ${\bf h}_S$ be the  $4 \times 1$ channel vector from an UE's antenna to all antennas of  the BS in the cell to which it belongs.  Let ${\bf h}_I$ be the $4 \times 1$ channel vector from an UE's antenna to all antennas of  any BS in the neighbouring cell. Signal to interference ratio (SIR) is defined as  ${\rm SIR} = \frac{{\bf h}_S^H{\bf h}_S}{{\bf h}_I^H{\bf h}_I}$. The various channel vectors are all generated such that ${\bf h}_S^H{\bf h}_S=1$ and ${\bf h}_I^H{\bf h}_I$ is such that the SIR is satisfied. 

\begin{figure}[t]
     \centering 
      \vspace {0.0in} 
        \includegraphics [width = 10cm,height = 8cm] {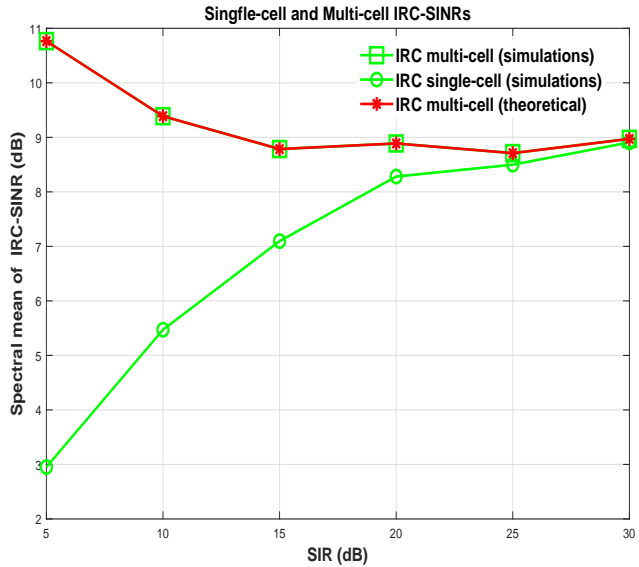}
         \caption{Comparisons between single-cell and multi-cell IRC-SINRs. $\sigma^2=0.1$. Results correspond to  25 iterations.}\label{one}
         \vspace{0.0in} 
\end{figure}

We define the spectral mean of a set of SNRs as follows. Suppose there are $A$ SNRs, ${\rm SNR}_1, \ldots, {\rm SNR}_A$, the spectral mean (SM) SNR of these SNRs is given by 
\begin{equation}
{\rm SNR}_{\rm SM} = \sqrt[A]{(1+{\rm SNR}_1)\ldots (1+{\rm SNR}_A)}-1\cdot
\end{equation}
The ${\rm SNR}_{\rm SM} $ of a set of $A$ SNRs is the SNR which will give the same spectral efficiency (as the sum of spectral efficiencies of corresponding to the $A$ SNRs) if each of the $A$ SNRs is replaced by ${\rm SNR}_{\rm SM} $. From Fig. \ref{one}, we can see the gain of IRC-SINR in the case of multi-cell scenario compared to the single-cell scenario. The gain is significant  at low SIRs where the interference is high. The ``multi-cell(theoretical)" curve in Fig. \ref{one} is obtained by starting with the ``single-cell (simulations)" curve in Fig. \ref{one} and applying \eqref{sims} with the values $N_R=4,a=12$. We see that ``multi-cell(theoretical)" and ``multi-cell(simulations)" curves overlap, thereby validating \eqref{theoretical}.

Now we give a numerical example to verify \eqref{theoretical}. Let $N_R=4, \sigma^2=0.1$, 
%

\begin{equation}
\begin{array}{*{20}c}
   {{\bf h}_1=\left[ {\begin{array}{*{20}c}
   {0.0841 + 0.0833i}  \\
{-0.2455 - 0.0302i  }   \\
{-0.5794 + 0.5822i  }   \\
{0.3141 + 0.3893i   }   \\
{0.0808 - 0.1263i  }   \\
 \end{array} } \right]}, & {{\bf P}=\left[ {\begin{array}{*{20}c}
      {0.0896 + 0.4466i} &  {-0.2823 + 0.0291i}  &  {-0.0967 + 0.1620i} \\
   { 0.2063 - 0.0202i } &   {  0.0948 - 0.2504i } &   { -0.2243 - 0.1287i}  \\
   {-0.0261 + 0.1448i  } &   { 0.3144 - 0.2070i   } &   {0.2673 - 0.1650i}  \\
   {0.1745 - 0.1172i } &   { -0.1434 - 0.0410i  } &   {-0.2230 + 0.2557i}  \\
   {-0.0984 - 0.2849i } &   { -0.0457 + 0.3269i   } &   {0.0004 + 0.3256i}  \\
 \end{array} } \right]}  \\

 \end{array} \cdot
\end{equation}
The values of $\text{IRC-SINR}^{\left( {N_R } \right)}, \text{IRC-SINR}^{\left( {N_R +1} \right)}$ in \eqref{snrgain3} are  5.8966 and 5.3994. respectively. The gain is $\text{IRC-SINR}^{\left( {N_R +1} \right)}- \text{IRC-SINR}^{\left( {N_R } \right)}= 0.4972$. The theoretical gain as per \eqref{theoretical} is indeed $\xi(N_R+1, N_R) = 0.4972$.
\section{Conclusions}
We derived and verified a theoretical analysis that gives the increase in IRC-SINR when the number of antennas increases by unity. We showed that this increase in IRC-SINR is always greater than or equal to zero.  Simulation in the context of uplink CoMP was carried out and verified with the theoretical analysis derived in this paper.

\end{document}